\documentclass[%
 reprint,
superscriptaddress,
 amsmath,amssymb,
 aps,prl
]{revtex4-1}

\usepackage{graphicx}% Include figure files
\usepackage{dcolumn}% Align table columns on decimal point
\usepackage{bm}% bold math
\usepackage[hidelinks]{hyperref}% add hypertext capabilities
\usepackage[dvipsnames]{xcolor}
\usepackage{comment}
\PassOptionsToPackage{caption=false}{subfig}
\usepackage{subfig}
\usepackage{multirow}
\usepackage{booktabs}
\usepackage{soul} 

\newcommand{\betahat}{\hat{\vect{\beta}}}
\newcommand{\vhat}{\hat{\vect{v}}}
\newcommand{\dd}{{\rm d}}

\newcommand{\vect}[1]{\boldsymbol{\mathbf{#1}}}

\begin{document}

%\preprint{APS/123-QED}

\title{First Results on Dark Matter Substructure from Astrometric Weak Lensing}

\author{Cristina Mondino}
 \email{cm4001@nyu.edu}
 \affiliation{Center for Cosmology and Particle Physics, Department of Physics, New York University, New York, NY 10003, USA}
\author{Anna-Maria Taki}
 \email{ataki@uoregon.edu}
\affiliation{Institute for Fundamental Science, Department of Physics, University of Oregon, Eugene, OR 97403, USA}
 \affiliation{Center for Cosmology and Particle Physics, Department of Physics, New York University, New York, NY 10003, USA}
\author{Ken Van Tilburg}
 \email{kvt@kitp.ucsb.edu}
 \affiliation{Kavli Institute for Theoretical Physics, University of California, Santa Barbara, CA 93106, USA}
 \affiliation{Center for Cosmology and Particle Physics, Department of Physics, New York University, New York, NY 10003, USA}
 \affiliation{School of Natural Sciences, Institute for Advanced Study, Princeton, NJ 08540, USA }
\author{Neal Weiner}
 \email{neal.weiner@nyu.edu}
 \affiliation{Center for Cosmology and Particle Physics, Department of Physics, New York University, New York, NY 10003, USA}

\date{\today}

\begin{abstract}
Low-mass structures of dark matter (DM) are expected to be entirely devoid of light-emitting regions and baryons. Precisely because of this lack of baryonic feedback, small-scale substructures of the Milky Way are a relatively pristine testing ground for discovering aspects of DM microphysics and primordial fluctuations on subgalactic scales. In this work, we report results from the first search for Galactic DM subhalos with time-domain astrometric weak gravitational lensing. The analysis is based on a matched-filter template of local lensing corrections to the proper motion of stars in the Magellanic Clouds. We describe a data analysis pipeline detailing sample selection, background subtraction, and handling outliers and other systematics. For tentative candidate lenses, we identify a signature based on an anomalous parallax template that can unequivocally confirm the presence of a DM lens, opening up prospects for robust discovery potential with full time-series data. We present our constraints on substructure fraction $f_l \lesssim 5$ at 90\% CL (and $f_l \lesssim 2$ at 50\% CL) for compact lenses with radii $r_l < 1\,\mathrm{pc}$, with best sensitivity reached for lens masses $M_l$ around $10^7$--$10^8\,M_\odot$. Parametric improvements are expected with future astrometric data sets; by end of mission, \textit{Gaia} could reach $f_l \lesssim 10^{-3}$ for these massive point-like objects, and be sensitive to lighter and/or more extended subhalos for $\mathcal{O}(1)$ substructure fractions.
\end{abstract}

\maketitle

%\tableofcontents

\section{Introduction}
%%%%%%%%%%%%%%%%%%%%%%%%%%%%%%%%%%%%%%%%%%%%%%%%%%%%%%%%%%%%%%%%%%%%%%%%%%%%%%%%%%%
The precise nature of the constituents of the dark matter (DM) and their microphysical properties is not known. Nevertheless, a wealth of information has been collected about its macroscopic properties and behavior from its minimal coupling to gravity and its resulting gravitational influence. In the linear theory of structure formation, a fluid with adiabatic fluctuations and vanishing sound speed clusters in a way that shows remarkable agreement with observations of the cosmic microwave background~\cite{ade2016planck,aghanim2018planck}, the Lyman-$\alpha$ forest~\cite{seljak2006cosmological,blomqvist2019baryon,agathe2019baryon}, and large-scale structures~\cite{percival2009testing,howlett2015clustering,zarrouk2018clustering}. This evolution has been probed over comoving length scales between $0.1\,\mathrm{Mpc}$ and $10^4\,\mathrm{Mpc}$ starting from a time when the Universe and the DM were more than 10 orders of magnitude denser than at the present time. N-body simulations~\cite{springel2008aquarius,diemand2008clumps,boylan2009resolving,stadel2009quantifying,garrison2014elvis,vogelsberger2014introducing,hellwing2016copernicus} extend these predictions to smaller physical length scales and into the nonlinear regime, matching observations of galactic rotation curves~\cite{rubin1970rotation, freeman1970disks, rogstad1972gross, whitehurst1972high, roberts1973comparison}, statistical dispersion relations~\cite{zwicky1933rotverschiebung,smith1936mass,zwicky1937masses, girardi1998optical, rines2006cirs, becker2007mean}, weak~\cite{kaiser1993mapping, schneider1996detection, wittman2000detection, hoekstra2004properties, okabe2014subaru} and strong~\cite{keeton2001cold, treu2010strong, jullo2010cosmological} gravitational lensing, and the distribution and abundance of satellite galaxies~\cite{klypin1999missing, willman2004observed, tollerud2008hundreds}. These complementary bodies of evidence not only put the existence of DM on a strong footing, they also provide knowledge about its coarse-grained phase-space distribution, indispensable input to searches for any nonminimal DM couplings. 

Detection of smaller DM structures becomes increasingly challenging due to their lower light-to-mass ratios; DM halos with scale masses below $10^8\,M_\odot$ do not harbor conditions for star formation and are thus entirely dark~\cite{rees1977cooling, kravtsov2010dark, bromm2013formation}. Methods reliant on the minimal coupling to gravity include fluctuations in extragalactic strong gravitational lenses~\cite{mao1998evidence,metcalf2001compound,chiba2002probing, dalal2002direct,metcalf2002flux,koopmans20022016,kochanek2004tests,inoue2005three,inoue2005extended,koopmans2005gravitational,chen2007astrometric,williams2008lensed,more2009role,keeton2009new,vegetti2009bayesian,vegetti2009statistics,congdon2010identifying,hezaveh2013dark,vegetti2014density,hezaveh2016measuring,hezaveh2016detection}, stellar wakes in the MW disk~\cite{feldmann2014detecting} and halo~\cite{buschmann2017stellar}, diffraction of gravitational waves~\cite{dai2018detecting}, photometric irregularities of micro-caustic light curves~\cite{dai2019gravitational}, and perturbations of cold stellar streams~\cite{ibata2002uncovering,johnston2002lumpy,siegal2008signatures,bovy2016detecting, carlberg2016modeling,erkal2016number,bonaca2018information} (with tentative positive detections~\cite{bonaca2018spur,Banik:2019cza}). These techniques show significant promise but are indirect or applicable to extragalactic structures only. Direct searches for MW substructure have so far been confined to transients in photometric lensing~\cite{paczynski1986gravitational,alcock2000macho,tisserand2007limits,griest2014experimental,
niikura2017microlensing,zumalacarregui2018limits} and pulsar timing~\cite{siegel2007probing,seto2007searching,baghram2011prospects,kashiyama2012enhanced,clark2015investigating,schutz2017pulsar,dror2019pulsar}, which only produce detectable signals for ultracompact objects such as black holes but not for more extended structures such as DM halos that collapse after matter-radiation equality.

In this Letter, we present the first results of a qualitatively new class of searches for Galactic DM substructure using time-domain, astrometric, weak gravitational lensing. Ref.~\cite{van2018halometry} proposed several categories of observables to this effect, and forecasted their sensitivity on upcoming astrometric surveys. We employ a refined version of their ``local velocity template'' on a sample of Small and Large Magellanic Cloud (SMC and LMC) stars in \textit{Gaia}'s second data release (DR2). Our data analysis constitutes a robust, optimal, matched-filter-based search for local distortions of the proper motion field of background sources produced by the gravitational lensing of intervening foreground compact DM subhalos. We find no evidence of this effect, setting a constraint of $f_l \equiv \Omega_l / \Omega_\mathrm{DM} \lesssim 5$ at 90\% CL (and $f_l \lesssim 2$ at 50 \% CL) for $M_l \sim 10^8 \, M_\odot$ and $r_l \lesssim 1\,\mathrm{pc}$, where $f_l$ is the DM substructure fraction, and $M_l$ and $r_l$ are the mass and characteristic radius of the subhalos. Currently limited by statistical instrumental uncertainties, we expect the reach to the combination $f_l M_l^2 / r_l^3$ to improve as $\propto t_\mathrm{int}^{-9/2}$, with the integration time $t_\mathrm{int}$ set to increase fivefold by \textit{Gaia}'s end of mission.

A localized discovery of dark low-mass substructures with our technique, possible with future astrometric surveys, would be a watershed event. Because of the absence of baryonic feedback, their abundance, mass function, and density profiles would provide a transparent window on the primordial fluctuation spectrum and the DM's transfer function on comoving scales below $0.1\,\mathrm{Mpc}$. It would probe the spectrum of adiabatic perturbations produced from the inflationary stage after the one measured in the CMB~\cite{ade2016planckinflation,akrami2018planck} and the Ly-$\alpha$ forest~\cite{bird2011minimally}, and of small-scale isocurvature fluctuations produced from e.g.~a late phase transition in the DM sector~\cite{zurek2007astrophysical,buschmann2019early}. Their discovery (non-observation) would rule out (provide evidence for) small-scale structure suppression, unavoidable predictions of light fermion (``warm'')~\cite{colin2000substructure, bode2001halo, viel2005constraining} and ultralight scalar (``fuzzy'')~\cite{hu2000fuzzy, li2014cosmological, hui2017ultralight} DM models. Enhanced-density subhalos can result from dissipation and self-interactions in the DM sector~\cite{Agrawal:2017pnb, Chang:2018bgx, essig2019constraining}, or early-time structure growth in axion DM models with large misalignment~\cite{arvanitaki2019large}.

\section{Lensing signal} 
\label{sec:lens_signal}
\begin{figure}
\centering
\includegraphics[width=0.32\textwidth]{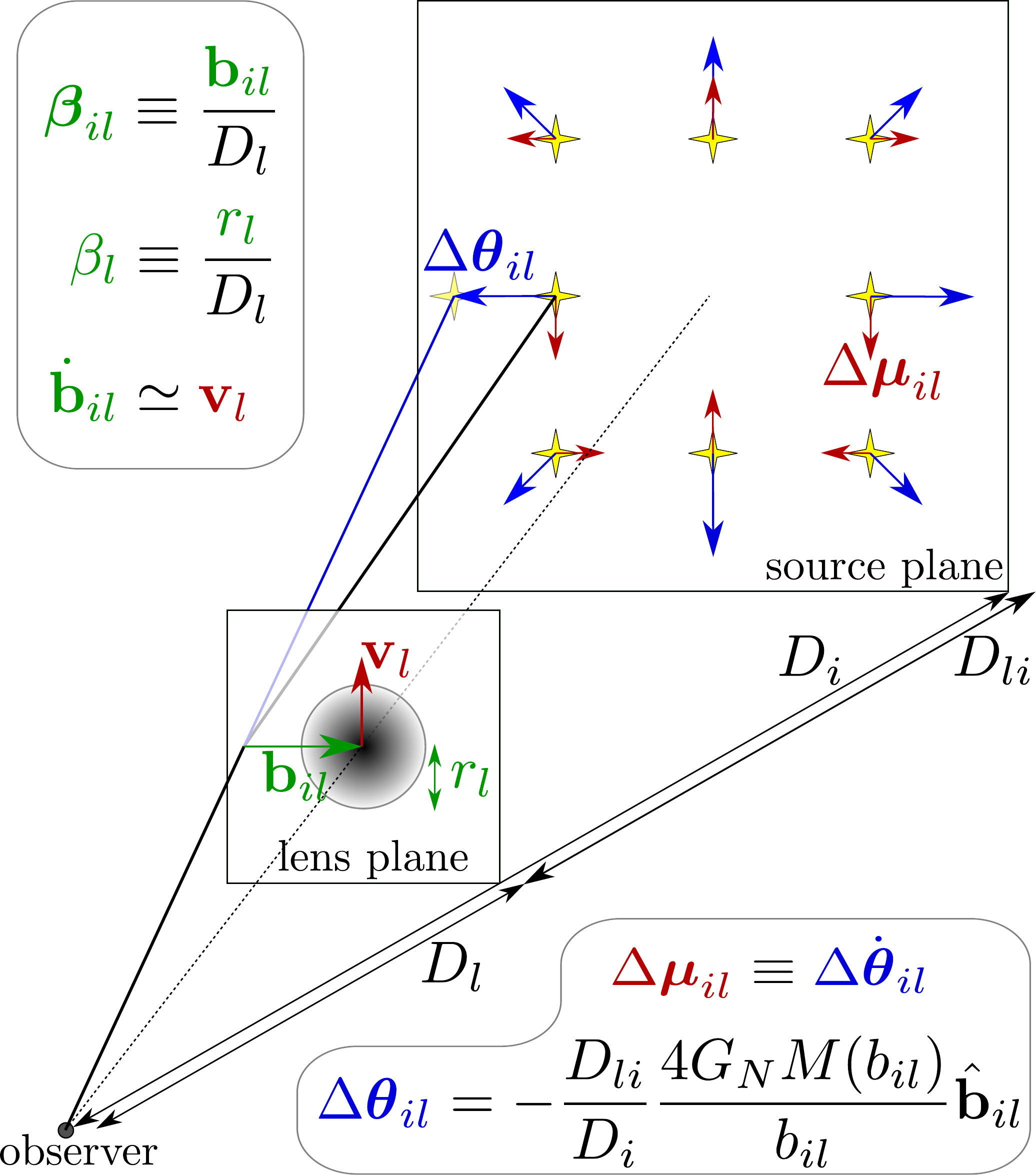}
\caption{Diagram of gravitational lensing geometry of sources $i$ by a lens $l$. The impact parameter is $\vect{b}_{il}$, its rate of change $\vect{v}_l$, the lens radius $r_l$, and respective line-of-sight distances $D_{i,l}$. In celestial coordinates, the angular impact parameter is $\vect{\beta}_{il} \equiv \vect{\theta}_i - \vect{\theta}_l$, angular lens radius $\beta_l$. The angular displacement $\Delta \vect{\theta}_{il}$ (blue, monopole pattern) is not constant in time, leading to lensing corrections $\Delta \vect{\mu}_{il}$ to the sources' proper motions $\vect{\mu}_i$ (red, dipole pattern).} \label{fig:diagram}
\end{figure}

The physical effect under consideration is time-domain weak gravitational lensing of the astrometric kind, summarized in Fig.~\ref{fig:diagram}. ``Weak'' refers to the regime where the impact parameter is much larger than the Einstein radius of the lens and one image of the source is resolved by the observer, and ``astrometric'' refers to the effect of angular deflection of the source's light centroid. True celestial positions $\vect{\theta}_i$ are unknown a priori, rendering the angular deflection $\Delta \vect{\theta}_{il}$ of source $i$ by lens $l$ unobservable in practice. 

Ref.~\cite{van2018halometry} proposed leveraging \emph{time-domain} lensing effects due to the relative rate of change in impact parameter $\dot{\vect{b}}_{il} \simeq \vect{v}_l$ in the reference frame of the observer. (We ignore the small contribution of the distant background source motions to $\dot{\vect{b}}_{il}$.) The leading observable in the time-domain (i.e.~to first order in $\vect{v}_l$) is a lensing correction to the proper motion $\vect{\mu}_i$:
\begin{align}
\Delta \vect{\mu}_{il} \equiv \Delta \dot{\vect{\theta}}_{il} =  \frac{D_{il}}{D_i} \frac{4 G_N M_l v_{l}}{r_{l}^2} \tilde{\vect{\mu}}_i(\beta_l, \vect{\beta}_{il}, \vhat_{l}), \label{eq:mumag}
\end{align}
where $G_N$ is Newton's gravitational constant, $r_l$ a characteristic lens radius, and $M_l = 4\pi \int_0^\infty \dd r \, r^2 \, \rho_l(r)$ the mass of the lens with 3D density profile $\rho_l(r)$. The unit-less 2D spatial profile of the distortion is 
\begin{align}
\tilde{\vect{\mu}}_i(\beta_l, \vect{\beta}_{il}, \vhat_{l}) &= \frac{\widetilde{M}_l(\beta_{il})}{\beta_{il}^2 / \beta_l^2} \left[2 \betahat_{il} (\betahat_{il}\cdot \vhat_{l}) - \vhat_{l}\right] \nonumber \\
&\phantom{=} - \frac{\partial_{\beta_{il}} \widetilde{M}(\beta_{il})}{\beta_{il}/\beta_l^2} \betahat_{il} (\betahat_{il}\cdot \vhat_{l}) \label{eq:mupattern}
\end{align}
with $M(b) = 2\pi \int_{-\infty}^{\infty} \dd z \, \int_0^{b} \dd b' \, b' \rho_l(\sqrt{z^2+b^{\prime 2}}) $ the enclosed lens mass within a cylinder oriented along the line of sight ($z$-direction) with radius equal to $b$, cfr.~Fig.~\ref{fig:diagram}, and $\widetilde{M}(\beta_{il}) \equiv M(b_{il})/M_l$. We also introduced the lens angular size $\beta_l \equiv r_l/D_l$ and angular impact parameter $\vect{\beta}_{il} \equiv \vect{b}_{il}/D_l$.

\begin{figure}[tbp]
	\includegraphics[width=0.48\textwidth]{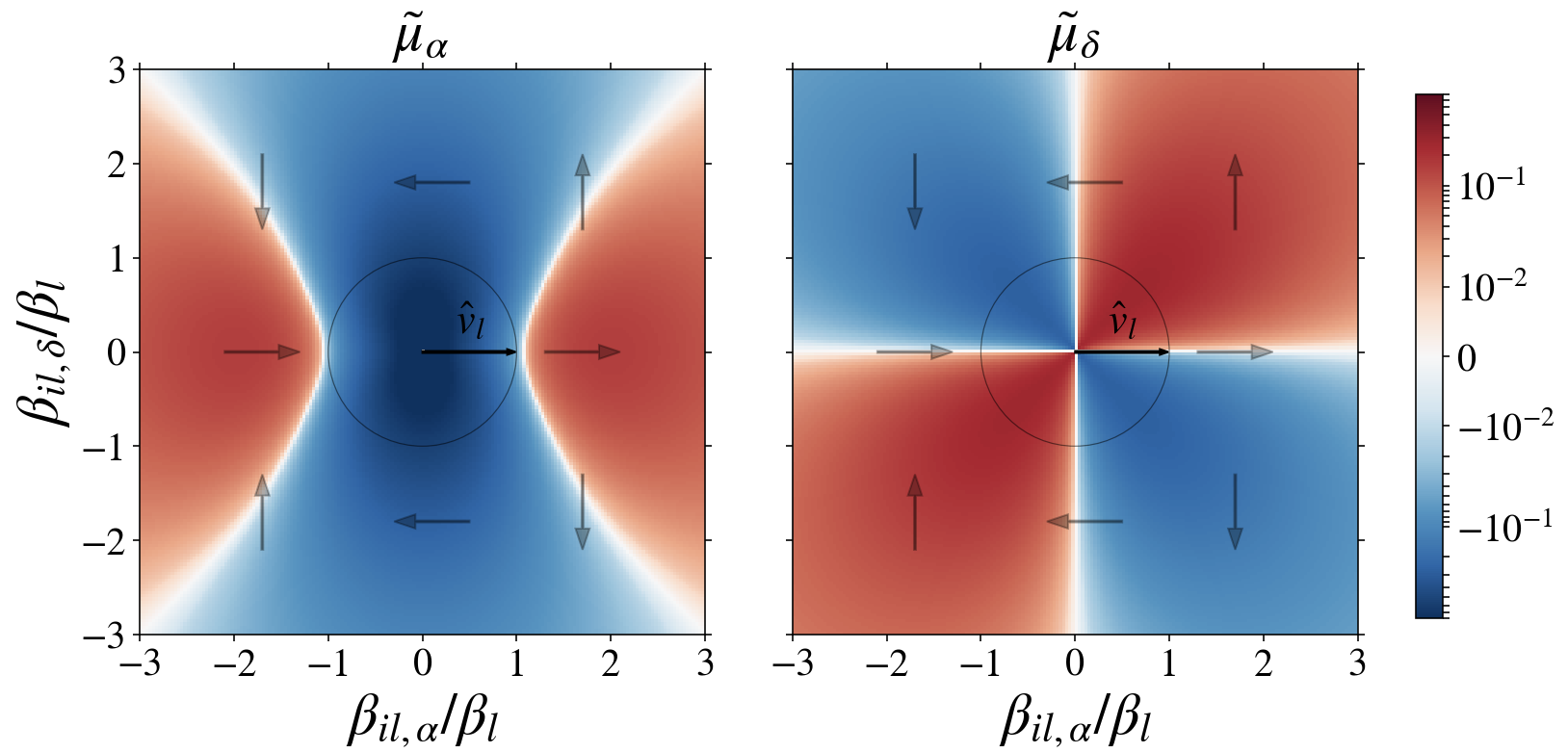}
	\caption{Right-ascension ($\alpha$, left) and declination ($\delta$, right) components of the angular velocity vector profile $\tilde{\vect{\mu}}_i$ of Eq.~\ref{eq:mupattern} as a function of angular separation $\vect{\beta}_{il}$ for the lens density profile of Eq.~\ref{eq:lensprofile}. The lens has angular size $\beta_l$ and is moving in the direction $\vhat_l = \vhat_\alpha$.}\label{fig:dipole}
\end{figure}

The primary lensing signature is thus a distortion in the angular velocity of background sources with a magnitude given by the prefactor in Eq.~\ref{eq:mumag} and the characteristic spatial pattern of Eq.~\ref{eq:mupattern}, which is a \emph{universal} dipole pattern for sources far outside the lens radius $\beta_{il}/\beta_l$ and depends on the lens density profile for sources eclipsed by the lens. For specificity, we will assume a lens density profile of
\begin{align}
\rho_l(r) = \frac{M_l}{4 \pi} \frac{\exp\lbrace - \frac{r^2}{2 r_l^2} \rbrace}{r r_l^2} \label{eq:lensprofile}
\end{align}
throughout the main text. The above profile exhibits the $1/r$ cusp of the NFW profile, but is nearly optimal in that it has no significant mass outside the scale radius (the radius where $\partial_{\ln r}( \ln  \rho_l) = -2$, in this case $r = r_l$), which would contribute to the lens mass abundance but only minimally to the lensing signal. The analysis presented here can easily be adapted to other density profiles, such as a pure Gaussian or a tidally truncated NFW profile, by properly choosing the function $\widetilde{M}_l$ in Eq.~\ref{eq:mupattern}. In the Appendix, we will investigate a profile-agnostic approach by truncating the lensing signal at $r_l$. In Fig.~\ref{fig:dipole}, we display the angular velocity distortion pattern of Eq.~\ref{eq:mupattern} resulting from the density profile in Eq.~\ref{eq:lensprofile}.

We take the lenses' spatial distribution across the Milky Way to follow that of the Galactic DM halo with a fiducial NFW profile 
\begin{align}
\rho(\vect{r}) = \frac{4 \rho_s}{\frac{r}{R_s}\big[1+\frac{r}{R_s}\big]^2}, \; R_s = 18\,\mathrm{kpc}, \; \rho_s = 0.003 \, \frac{\mathrm{M_\odot}}{\mathrm{pc}^3}, \label{eq:den_profile}
\end{align}
with $r$ the galactocentric radius, and the observer located at $r_\odot = 8\,\mathrm{kpc}$ \cite{mcmillan2011mass}. We assume a transverse lens velocity distribution for the subhalos given by:
\begin{align}
	p_v(\vect{v}_l) = \frac{1}{2\pi\sigma_v^2} \exp\left[-\frac{\left(\vect{v}_l - \vect{v}_0\right)^2}{2\sigma_v^2}\right], \;  \sigma_v \simeq 166 \, \frac{\mathrm{km}}{\mathrm{s}}, \label{eq:vdistro}
\end{align}
where $\vect{v}_0$ denotes a 2D velocity vector equal in magnitude and opposite to the observer's velocity projected on the plane perpendicular to the line of sight. We take the observer's velocity to be the Solar System velocity of $238$ km/s in the Galactic equatorial plane ($b\approx 0$, $l \approx 270^\circ$). In the following, we will ignore the additional annual rotation around the Sun, but will return to this parallax effect in the Discussion.
%%%%%%%%%%%%%%%%%%%%%%%%%%%%%%%%%%%%%%%%%%%%%
\section{Template method}

%\paragraph{Local test statistic---}
We utilize a \emph{local} test statistic $\mathcal{T}$ that computes the overlap of the velocity field of background sources with the one induced by a tentative lens candidate with angular position $\vect{\theta}_t$, angular scale $\beta_t$, and effective lens velocity direction $\vhat_t$~\cite{van2018halometry}:

\begin{align}
	\mathcal{T}(\vect{\theta}_t,\beta_t, \vhat_t) &\equiv \sum_{i} \frac{\vect{\mu}_i \cdot  \tilde{\vect{\mu}}_i(\beta_t, \vect{\beta}_{it}, \vhat_t)}{\sigma_{\mu,i}^2}, \label{eq:template_stat}
\end{align}
where $\vect{\mu}_i\equiv \lbrace\mu_{i,\alpha *}, \mu_{i,\delta}\rbrace = \lbrace \mu_{i,\alpha} \cos{\delta_i}, \mu_{i,\delta}\rbrace$ is the proper motion vector of the $i$th star, and $\sigma_{\mu,i}^2 \equiv \sigma_{\mu_{\alpha},i}^2 + \sigma_{\mu_{\delta},i}^2$ is the measured variance over the chosen stellar population. The velocity template vector $\tilde{\vect{\mu}}_i$ is a matched filter to the lens-induced velocity vector profile and is given in Eq.~\ref{eq:mupattern} generally, and for the specific density profile of Eq.~\ref{eq:lensprofile} in Fig.~\ref{fig:dipole}. $\mathcal{T}$ depends on the lens position through the angular impact parameters $\vect{\beta}_{it} \equiv \vect{\theta}_t - \vect{\theta}_i$.

We define the normalization factor
\begin{align}
	\mathcal{N}^2(\vect{\theta}_t,\beta_t) &\equiv \sum_{i} \frac{|\tilde{\vect{\mu}}_i(\beta_t, \vect{\beta}_{it}, \vhat_t)|^2}{\sigma_{\mu,i}^2}, 
	 \label{eq:template_norm}
\end{align}

that acts as a figure of merit for the sensitivity of a candidate lens position and radius: large values indicate the presence of numerous low-noise stars within $\beta_t$ around the template. 
In absence of a lensing signal, one expects vanishing mean $\langle \mathcal{T} \rangle_\mathrm{n}$ with a variance $\langle \mathcal{T}^2 \rangle_\mathrm{n} = \mathcal{N}^2 \sim \Sigma \beta_t^2 / \sigma_{\mu}^2$ with $\Sigma$ the typical local angular number density of background sources.

In the presence of a lens, and with template parameters perfectly matched to those of the lens (i.e.~$\vect{\theta}_t = \vect{\theta}_l$, $\beta_t = \beta_l$, and $\hat{\vect{v}}_t = \hat{\vect{v}}_l$), the test statistic is expected to evaluate to $\langle \mathcal{T} \rangle_\mathrm{s} \simeq  \mathcal{N}^2 4 G_N M_l v_l/ r_l^2$ for a nearby lens $D_l \ll D_i$. The \emph{local} signal-to-noise ratio
\begin{align}
\mathrm{SNR} = \frac{\langle \mathcal{T} \rangle_\mathrm{s}}{\sqrt{\langle \mathcal{T}^2 \rangle_\mathrm{n}}} \simeq \frac{4 G_N M_l v_l}{r_l^2}\mathcal{N} \sim \frac{4G_N M_l v_l \sqrt{\Sigma}}{r_l D_l \sigma_\mu} \label{eq:SNRlocal}
\end{align}
is generally largest for the most nearby, massive, compact, and fast-moving lenses in front of high-density, low-noise regions.

The true lens properties are unknown a priori. We evaluate $\mathcal{T}$ over a dense grid in $\vect{\theta}_t$ and $\beta_t$, and for two lens velocity directions (along RA and DEC) 
$\mathcal{T}_{\alpha}(\vect{\theta}_t,\beta_t)  \equiv \mathcal{T}(\vect{\theta}_t,\beta_t, \hat{\vect{\alpha}})$ and similarly for $\mathcal{T}_\delta$ which we will combine into a vector $\vect{\mathcal{T}} \equiv \lbrace \mathcal{T}_\alpha, \mathcal{T}_\delta \rbrace$. The directional asymmetry in Eq.~\ref{eq:vdistro} translates into the same preferred direction for the template $\langle \vect{\mathcal{T}} \rangle_\mathrm{s} \propto \vect{v}_0$.

We define a \emph{global} test statistic $\mathcal{R}$ that is the optimal observable (see Appendix for derivation) for detecting the proper motion distortion of a \emph{single} lens across a certain patch of sky:
\begin{align}
\hspace{-0.85em}	\mathcal{R} = \sup_{\vect{\theta}_t, \beta_t} \Bigg [ \ln \frac{\rho}{\beta_t^4} + \frac{C^2 \sigma_v^2 \mathcal{N}^2\Big(\frac{\mathcal{T}^2}{\mathcal{N}^2}  -  \frac{v_0^2}{\sigma_v^2}  \Big)+ 2 C \vect{\mathcal{T}}	\cdot\vect{v}_0}{2(1+C^2 \sigma_v^2 \mathcal{N}^2)} \Bigg ]
	\label{eq:test_statistic_R}
\end{align}
with $\rho = \rho(\vect{\theta}_t,r_l/\beta_t)$ from Eq.~\ref{eq:den_profile}, $C = 4 G_N M_l/r_l^2$ and $\mathcal{T}^2 = \mathcal{T}_\alpha^2 +  \mathcal{T}_\delta^2$. Roughly speaking, it corresponds to taking the largest value of $\vect{\mathcal{T}}/\mathcal{N}$ across the densely-scanned grid of $\lbrace \vect{\theta}_t,\beta_t \rbrace$, but it also properly accounts for the $\vect{v}_l$ asymmetry, variations in $\mathcal{N}(\theta_t,\beta_t)$, and priors on the 3D location of the lens. 

\section{Data Processing}
\paragraph{Data sample---} For our analysis, we choose astrometric data on the Large and Small Magellanic Clouds (LMC and SMC) from \textit{Gaia}'s second data release~\cite{prusti2016gaia, brown2018gaia}. They have large stellar angular number densities and low proper motion dispersion (intrinsic and instrumental), maximizing the SNR of Eq.~\ref{eq:SNRlocal} with a high $\sqrt{\Sigma}/\sigma_\mu$. Their large combined angular area also increases the probability of at least one nearby (low $D_l$) lens. 

To avoid foreground contamination, we select sources without evidence of parallax ($\varpi$ /$\sigma_\varpi < 2$) in a square of $10^\circ$ sidelength centered on $(\alpha, \delta)=(78.77^\circ, -69.01^\circ)$ for the LMC and $8^\circ$ on $(12.80^\circ, -73.15^\circ)$ for the SMC. For the SMC, we impose $|\mu_{\alpha*} - 0.685\,\mathrm{mas/y}| < 2\,\mathrm{mas/y}$ and $|\mu_\delta + 1.230\,\mathrm{mas/y}| < 2\,\mathrm{mas/y}$ to cut out the foreground NGC 104 and NGC 362 globular clusters.
Poor astrometric solutions were avoided with a cut on Renormalized Unit Weight Error of $\mathrm{RUWE} < 1.4$~\cite{lindegren2018re}. We summarize our data processing operations here and refer the interested reader to the Appendix for more details. 

\paragraph{Removal of dense clusters---} Overdense stellar clusters generally move coherently and independently from the bulk stars in the Magellanic Clouds (MCs), and are thus contaminants from our perspective. 
We calculate a smoothed angular number density map $\Sigma_\mathrm{sm}(\vect{\theta})$ with a Gaussian kernel of angular radius $0.1^\circ$ and  a pixelated angular number density map $\Sigma(\vect{\theta})$ with pixels of size $0.1^\circ /3$. Density outliers are removed by excising regions for which $\Sigma> 3 \Sigma_\mathrm{sm}$. 

\paragraph{Motion subtraction and outlier removal---} We subtract the large-scale proper motion and remove stars that are not bound to the MCs. Operationally, we define a motion field $\vect{\mu}(\vect{\theta}_p) = \sum_{i \in p} \vect{\mu}_i \sigma_{\mu,i}^{-2} / \sum_{i \in p} \sigma_{\mu,i}^{-2} $ in square pixels $p$ of $0.05^\circ$, from which we calculate a smoothed motion field $\vect{\mu}_\mathrm{sm}(\vect{\theta})$ with Gaussian kernel of radius $0.1^\circ$. We then construct a list of stellar motions with large-scale motion subtracted: $\vect{\mu}_{\mathrm{sub},i} \equiv \vect{\mu}_i - \vect{\mu}_\mathrm{sm}(\vect{\theta}_i)$. Stars with $\mu_{\mathrm{sub},i} > 3 \sigma_{\mu,i} + \mu_\mathrm{esc}$ are removed, were $\mu_\mathrm{esc}$ is the (proper) escape velocity. The outlier removal slightly biases $\vect{\mu}_\mathrm{sm}$, so the process is iterated another two times with the remaining stars. 

\begin{center}
\begin{figure}[tbp]
	\includegraphics[width=0.45\textwidth]{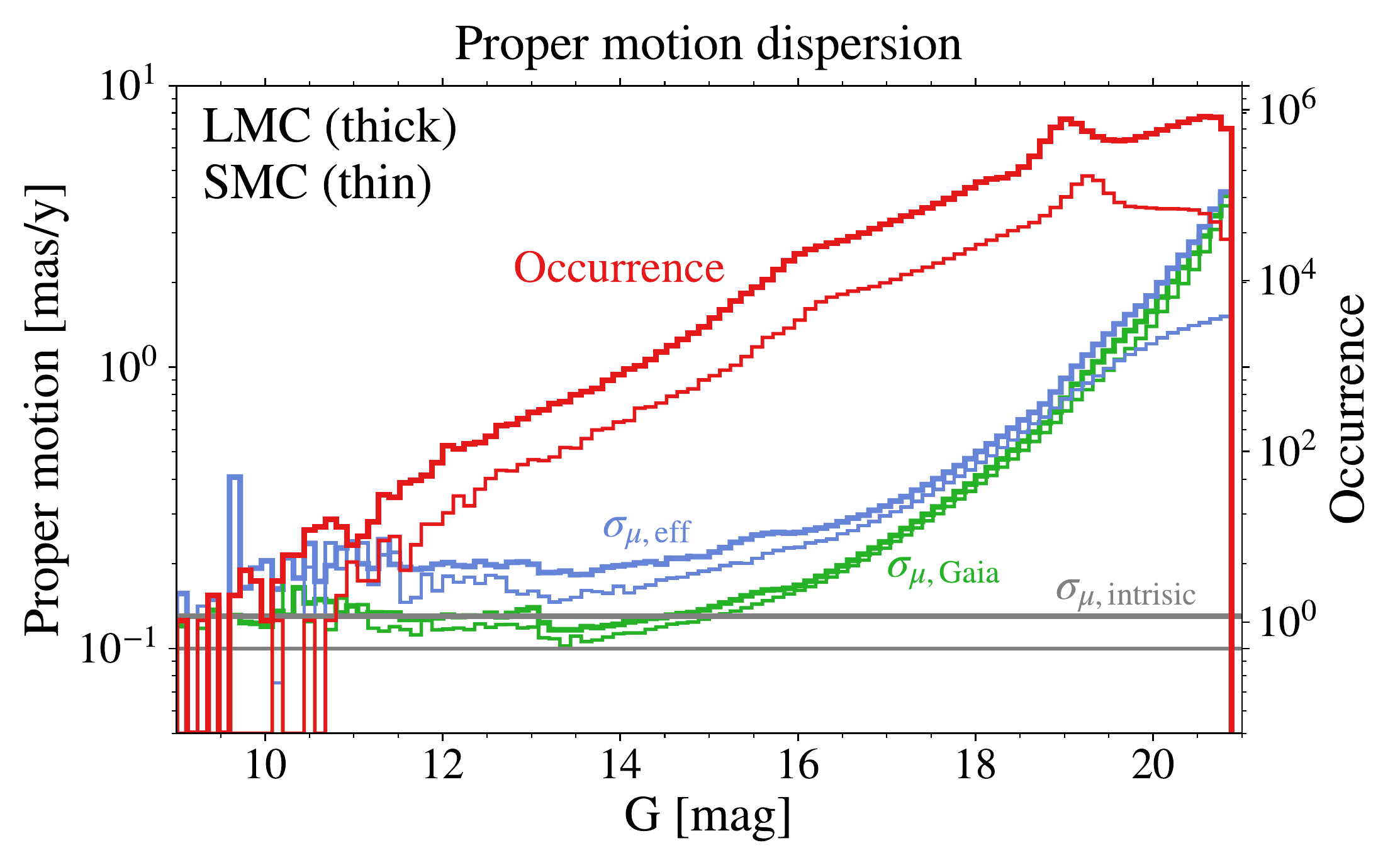}
	\caption{Number of stars (red), observed ($\sigma_{\mu,\text{eff}}$, blue), reported ($\sigma_{\mu,\textit{Gaia}}$, green), and intrinsic ($\sigma_{\mu,\text{intrinsic}}$, gray) proper motion dispersion as a function of stellar G magnitude for the LMC (thick) and SMC (thin). }\label{fig:pm_disp}
\end{figure}
\end{center}

\paragraph{Effective error---} After the above procedures, we arrive at a proper motion field with $\langle \vect{\mu}_{\mathrm{sub},i} \rangle \simeq 0$ but where the observed variance $\sigma_{\mu,\mathrm{eff}}^2 \equiv \langle \mu_{\mathrm{sub},i}^2 \rangle$ still exceeds the \textit{Gaia}-reported variance $\sigma_{\mu,\mathrm{Gaia}}^2 \equiv \langle \sigma_{\mu,i}^2 \rangle$. This discrepancy is due to intrinsic (proper) velocity dispersion $\sigma_{\mu,\mathrm{intrinsic}}$ in the MCs as well as unmodeled instrumental systematics, unresolved binaries and double stars, and other astrometric misfits \cite{arenou2018gaia, lindegren2018gaia}. In Fig.~\ref{fig:pm_disp}, we plot the number of stars (red), $\sigma_{\mu,\mathrm{eff}}$ (blue), $\sigma_{\mu,\mathrm{Gaia}}$ (green), and $\sigma_{\mu,\mathrm{intrinsic}}$ (gray) \cite{gyuk2000self,evans2008kinematics} in bins of 0.1 width in G magnitude for the LMC (thick) and SMC (thin). In the following analysis, we use the G-mag-dependent $\sigma_{\mu,\mathrm{eff}}^2$ as the inverse weight factor in Eqs.~\ref{eq:template_stat} and \ref{eq:template_norm}.

\section{Analysis \& Results}
\begin{figure}[tbp]
	\includegraphics[width=0.45\textwidth]{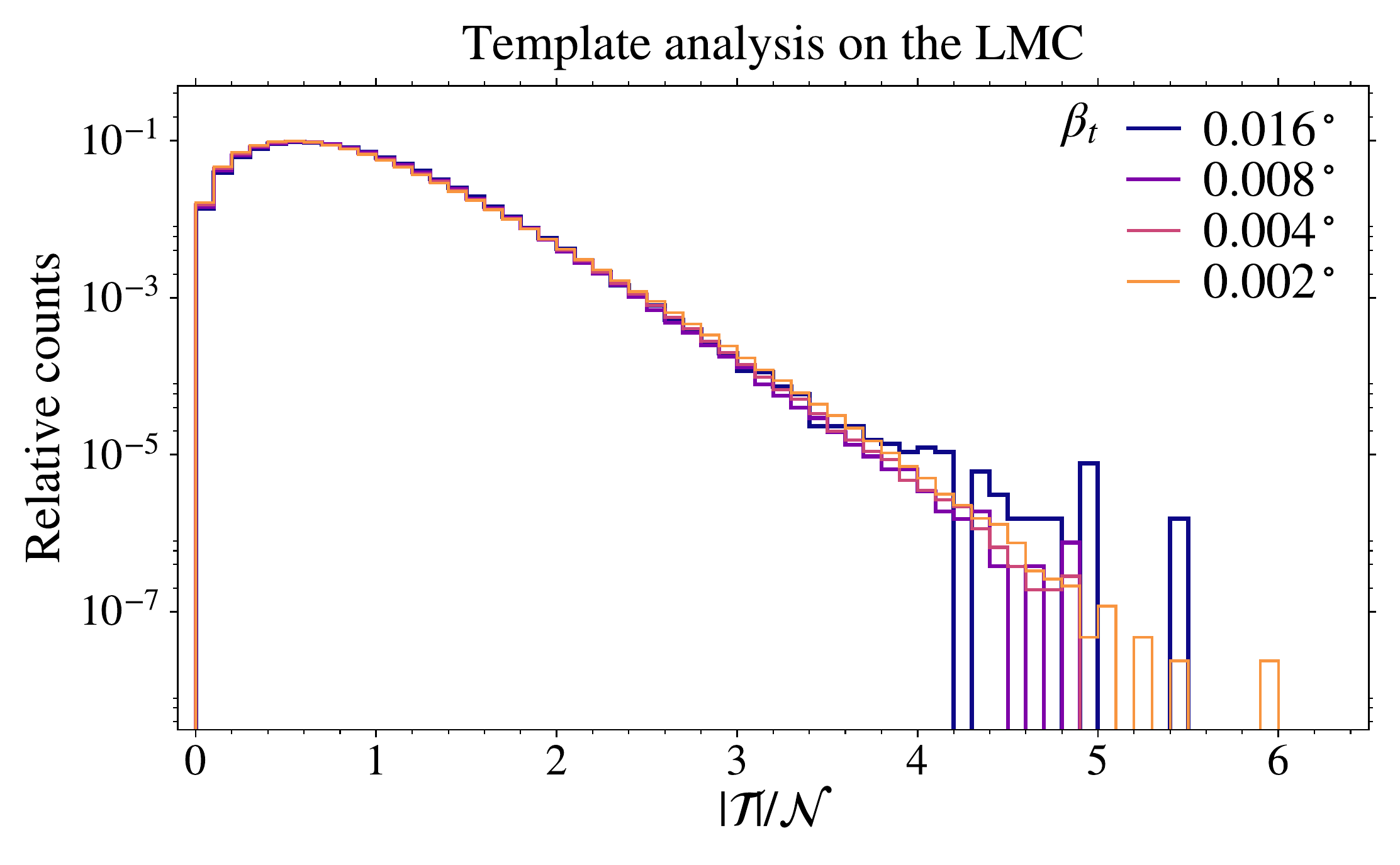}
	\caption{Histograms of the local test statistic $\mathcal{T}(\vect{\theta}_t,\beta_t)/\mathcal{N}(\vect{\theta}_t,\beta_t)$, where $\mathcal{T} = \sqrt{ \mathcal{T}_\alpha^2+ \mathcal{T}_\delta^2}$, evaluated at all template locations $\vect{\theta}_t$ and different angular template radii $\beta_t$ for the coarse-grid analysis on the LMC data sample.}\label{fig:Tad}
\end{figure}

\paragraph{Evaluation of test statistics---} We compute $\vect{\mathcal{T}}(\vect{\theta}_t,\beta_t)$ and $\mathcal{N}(\vect{\theta}_t,\beta_t)$ over a coarse square grid with lattice constant $\beta_\mathrm{scan} = 0.9 \times \beta_t $ for a fixed list of 58 $\beta_t$ values evenly spaced between $0.0015^\circ$ and $0.03^\circ$. The results of this procedure for 4 angular scales are displayed for the LMC data in Fig.~\ref{fig:Tad}, and exhibit a near-Gaussian distribution of $\mathcal{T}/\mathcal{N}$ out to $5$ sigma. 
At each $\beta_t$, we then identify coarse lattice sites at which $\mathcal{T} > 0.25 \max_{\vect{\theta_t}}\mathcal{T}$, and compute $\vect{\mathcal{T}}, \mathcal{N}$ over sets of finer grids with lattice constant $\beta'_\mathrm{scan} = \beta_\mathrm{scan}/3$ around these high-$\mathcal{T}$ sites. This finer scanning procedure is iterated once more with $\beta''_\mathrm{scan} = \beta'_\mathrm{scan}/3$ at lattice sites at which $\mathcal{T} > 0.5 \max_{\vect{\theta_t}}\mathcal{T}$.
For each parameter space point in the 3D space $\lbrace M_l, r_l, f_l \rbrace$, we calculate $\mathcal{R}$ as defined in Eq.~\ref{eq:test_statistic_R} from the finest grid of $\vect{\mathcal{T}}, \mathcal{N}$ values over the combined LMC/SMC sample.

\paragraph{Signal simulations---} For each point in $\lbrace M_l, r_l, f_l \rbrace$, we create a minimum of 100 simulations of lensing signal and stochastic noise. In each simulation, we generate a random number $N_l$ of lenses from a Poisson distribution with mean $\langle N_l \rangle = \Delta \Theta\ (\Omega_l / M_l)  \int_0^{D_i} \dd D_l \, D_l^2 \rho(\vect{\theta}_i, D_l)$ with $\rho(\vect{r})$ from Eq.~\ref{eq:den_profile} and $\Delta \Theta$ the solid angle subtended by the data, for both the LMC and SMC. The pdf for the 3D position of the lenses $(D_l, \vect{\theta}_l)$ (determining $\beta_l$) is taken proportional to $\rho(\vect{r})$, and that for $\vect{v}_l$ is given by Eq.~\ref{eq:vdistro}. 

In each of the simulations, we inject stochastic proper motion noise. We first group the stars in two-dimensional bins of $0.05$ width in G magnitude and $1^\circ$ radial bins from the centers of the LMC and SMC, and then deduce the proper motion pdf by equating it to the observed distribution of $\vect{\mu}_{\mathrm{sub},i}$ in each bin on the data samples. (These pdfs are decidedly non-Gaussian, only the variance of their 1D G-mag projection is shown in Fig.~\ref{fig:pm_disp}). Finally, we produce signal-plus-noise simulations by random draws from these proper motion noise pdfs, and by subsequent distortions via Eq.~\ref{eq:mumag} from their associated random lens population. 

\paragraph{Constraints---} The simulations are run through the exact same data processing and analysis pipeline (in particular also the motion subtraction and outlier removal) as the actual data, yielding a distribution of $\mathcal{R}$ values for each parameter space point in $\lbrace M_l, r_l, f_l \rbrace$. If 90\% (50\%) of the simulations have an $\mathcal{R}$ value larger than the observed $\mathcal{R}$ value for any parameter space point, then that point is excluded at 90\% (50\%) CL. As a cross-check on our limit-setting strategy, we also generated 60 noise-only simulations, and found that the mean $\mathcal{R}$ value across a handful of parameter space points is 92\%--97\% that of the observed value. This observation implies that no significant excess is present in the data, and that our noise injection is conservative.

\begin{figure}[tbp]
	\includegraphics[width=0.45\textwidth]{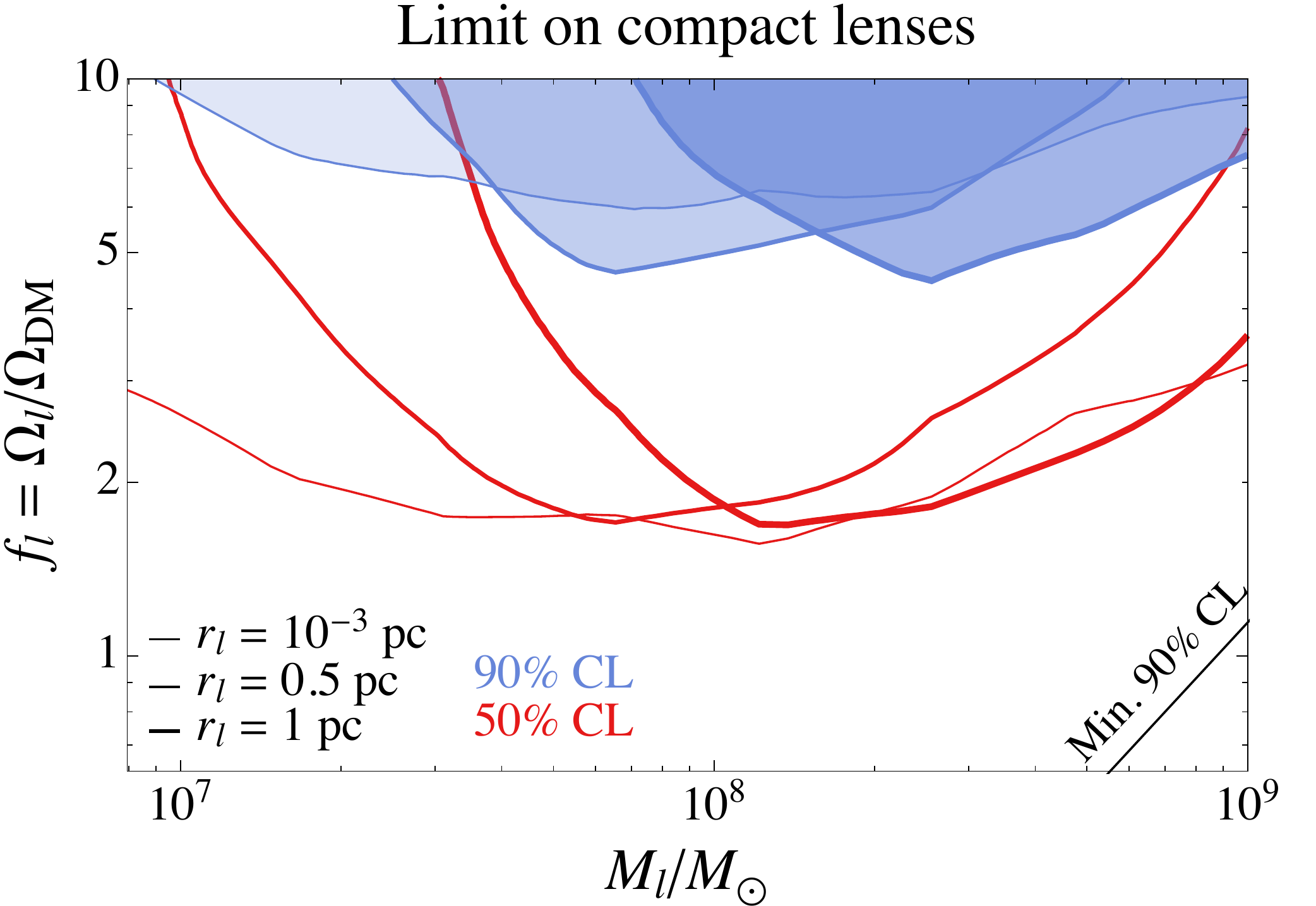}
\caption{Constraints from the MCs velocity template analysis on the fractional dark matter abundance $f_l$ of compact objects with mass $M_l$ and density profile given in Eq.~\ref{eq:lensprofile}, for different compact object radii $r_l = 10^{-3}, 0.5, \text{ and }1~\text{pc}$. The constraint for the smallest radius is equivalent to the one for point-like objects ($r_l = 2 G_N M_l$) given the angular number density of stars. Above the diagonal line at the bottom right, at least one subhalo eclipses the data sample with 90\% probability. }\label{fig:compactlimit}
\end{figure}

The resulting limits are displayed in the $(M_l, f_l)$ plane in Fig.~\ref{fig:compactlimit} for three values of $r_l$. The current data set is sensitive to substructure fractions $f_l$ between 1 and 5 from for $M_l$ between $10^6 \, M_\odot$ and $10^9 \, M_\odot$, while the 90\% CL limit reaches $f_l\approx 5$ only at the most sensitive parameter-space points. The comparatively worse limit for $r_l = 0.5\,\mathrm{pc}$ (relative to $r_l = 10^{-3}\,\mathrm{pc}, 1\,\mathrm{pc}$) at high $M_l$ is driven by a relatively high maximum $\mathcal{T}/\mathcal{N}$ value for $\beta_t = 0.002^\circ$ near $\vect{\theta}_t \approx (82.54^\circ,-69.76^\circ)$, consistent with a statistical fluctuation for a background-only hypothesis.

%%%%%%%%%%%%%%%%%%%%%%%%%%%%%%%%%%%%%%%%%%%%%%%%%%%%%%%%%%%%%%%%%%%%%%%%%%%%%%%%%%%
\section{Discussion} 
We presented results and limits from the first analysis leveraging precision astrometry data and time-domain weak gravitational lensing to look for Galactic substructure. A simple modification of our analysis technique can also unambigously \emph{confirm} tentative local lensing signals. The rate of change of impact parameter receives periodic contributions from Earth's motion around the Sun with known phase, direction, and magnitude of order $30\,\mathrm{km/s}$. This lensing-induced ``anomalous parallax'' motion is guaranteed to be present if a tentative signal of correlated linear stellar motions is due to astrometric weak lensing. It is almost an order of magnitude smaller, and its error has less favorable scaling with integration time than the linear motion ($\sigma_{\mu,\varpi}/\sigma_{\mu} \sim t_\mathrm{int} / \mathrm{1\,yr}$). However, this signature will always be statistics-limited insofar that it cannot be faked by intrinsic stellar motions. This latter observation opens up the possibility of astrometric lensing searches for dark matter on \textit{Gaia}'s \emph{entire} data set (once time-series data becomes available), including more precisely measured nearby stars rather than the distant MCs.

Our constraints are statistics-limited now and for the foreseeable future, as the figure-of-merit $\sqrt{\Sigma}/\sigma_{\mu,\mathrm{eff}}$ is largest for relatively faint stars with $\mathrm{G} > 16$ and the vast majority of MC stars have $\mathrm{G} > 19$. With integration time $t_\mathrm{int}$, currently at 22 months for \textit{Gaia} DR2, we expect the proper motion error to scale at least as fast as $\sigma_{\mu,\mathrm{eff}} \propto t_\mathrm{int}^{-3/2}$. The scaling will likely be faster as more stars are added, binaries and double stars are resolved, and modeling of telescope systematics improves with time. Note that Eq.~\ref{eq:SNRlocal} is valid for $r_l \gtrsim 1\,\mathrm{pc}$ in this context and that the closest lens at $\langle \min_l D_l \rangle \propto (M_l / f_l)^{1/3}$ drives the sensitivity. We project that the sensitivity to the combination $f_l M_l^2 / r_l^3$ will improve as $\sigma_{\mu, \mathrm{eff}}^3 \propto t_\mathrm{int}^{-9/2}$ or better, yielding promising prospects for future data releases from $\textit{Gaia}$ and other astrometric surveys.
\clearpage

\begin{acknowledgments}
We thank Asimina Arvanitaki, Masha Baryakhtar, David Hogg, Junwu Huang, Mariangela Lisanti, Siddharth Mishra-Sharma, and Oren Slone for helpful discussions.
%kvt funding
KVT's research is funded by the Gordon and Betty Moore Foundation through Grant GBMF7392.
%CM funding
CM is supported by the Thomas J. Moore dissertation fellowship.
%amt funding
AMT is supported in part by the U.S.~Department of Energy under Grant Number DE-SC0011640.
%NW funding
NW is funded by the Simons Foundation and by the NSF under Grant No. PHY-1620727 and PHY-1915409.
%gaia blurb
This project was developed in part at the April 2018 NYC Gaia DR2 Workshop and the 2018 NYC Gaia Sprint at the Center for Computational Astrophysics of the Flatiron Institute. This work has made use of data from the European Space Agency (ESA) mission {\it Gaia} (\url{https://www.cosmos.esa.int/gaia}), processed by the {\it Gaia} Data Processing and Analysis Consortium (DPAC, \url{https://www.cosmos.esa.int/web/gaia/dpac/consortium}). Funding for the DPAC has been provided by national institutions, in particular the institutions participating in the {\it Gaia} Multilateral Agreement.
\end{acknowledgments}

\appendix

\section{Appendix}
%\appendix

\subsection{Derivation of optimal discriminant}
\label{app:opt_discriminant}

We present the derivation of our likelihood-inspired test-statistic $\mathcal{R}$, a \emph{global} analog of the {\it local} test-statistic $\mathcal{T}$ that appropriately weighs over all possible lens locations, angular sizes and velocity directions. Ideally, we would like to check whether the stellar proper motions $\lbrace \vect{\mu}_i \rbrace$ observed across a certain patch of the sky are compatible with the proper motion distortions induced by a population of foreground lenses. The full likelihood function for a lens population is hardly tractable due to the large number of random variables involved. However, we can simplify it by including only the contribution from a {\it single} lens, noting that the local signal-to-noise ratio is driven by the closest one (see Eq.~\ref{eq:SNRlocal}). In addition, we regard the measured stellar proper motions as independent Gaussian random variables with zero (subtracted) mean. Within this approximation, the likelihood function for a {\it single} lens originating from a population of lenses with mass $M_l$, characteristic physical size $r_l$ and fractional abundance $f_l = \Omega_l / \Omega_\mathrm{DM}$ reads

\begin{align}
	\hspace{-0.5em}&\mathcal{L}\left( \lbrace \vect{\mu}_i \rbrace \big |M_l,r_l, f_l \right) = p_1(\vect{\theta}_t, \beta_t) p_v(\vect{v}_t)  \prod_i \frac{\exp\left[-\frac{(\vect{\mu}_i-\Delta\vect{\mu}_i)^2}{2\sigma_{\mu,i}^2} \right]}{2\pi \sigma_{\mu,i}^2} ,\label{eq:likelihoods}
\end{align}
where the lens correction $\Delta\vect{\mu}_i$ is given by Eq.~\ref{eq:mumag}, $p_v$ is the pdf for the tentative lens velocity $\vect{v}_t$ from Eq.~\ref{eq:vdistro}, and $p_1$ corresponds to a joint pdf for the tentative lens position $\vect{\theta}_t$ and size $\beta_t$
\begin{align}
	p_1(\vect{\theta}_t, \beta_t) = \frac{r_l^3 \rho(\vect{\theta}_t, r_l/\beta_t)}{\beta_t^4 M_l \langle N_l \rangle}, 
\end{align}
with $\rho(\vect{r})$ in Eq.~\ref{eq:den_profile}, and $\langle N_l \rangle$ the expected number of lenses in front of the stellar target. The log-likelihood ratio gives
\begin{widetext}
\begin{align}
	\ln \frac{ \mathcal{L}\left( \lbrace \vect{\mu}_i \rbrace \big |M_l,r_l, f_l \right)}{\mathcal{L}\left( \lbrace \vect{\mu}_i \rbrace \big |\text{no lens} \right)} &= \ln\frac{r_l^3}{M_l \langle N_l\rangle}+\ln \frac{\rho(\vect{\theta}_t, r_l/\beta_t)}{\beta_t^4} -\frac{\left(\vect{v}_t - \vect{v}_0\right)^2}{2\sigma_v^2} + C \vect{\mathcal{T}}\cdot \vect{v}_t  - \frac{C^2v_t^2\mathcal{N}^2 }{2},
	\label{eq:likratio}
\end{align}
\end{widetext}
where $C=4 G_N M_l/r_l^2$, and we have introduced $\vect{\mathcal{T}} \equiv \lbrace \mathcal{T}(\theta_t, \beta_t, \vect{\hat{\alpha}}), \mathcal{T}(\theta_t, \beta_t, \vect{\hat{\delta}}) \rbrace$, with $\mathcal{T}$ defined in Eq.~\ref{eq:template_stat}. The normalization factor is defined in Eq.~\ref{eq:template_norm}; in the limit of a large number of stars distributed in a circularly symmetric way around $\vect{\theta}_t$, it approaches:
\begin{align}
	\hspace{-0.85em}\mathcal{N}^2(\vect{\theta}_t,\beta_t) 
	                                                            \simeq \sum_{i} \frac{\beta_t^4/\beta_{it}^4}{\sigma_{\mu,i}^2}\left[\widetilde{M}_l^2 + \frac{(\partial \widetilde{M}_l)^2}{2\beta_t^2/\beta_{it}^2}-\frac{\widetilde{M}_l\partial\widetilde{M_l}}{\beta_t/\beta_{it}}\right]. \label{eq:template_norm_simp} 	                                          
\end{align}
The optimal test statistic is given by maximizing the likelihood ratio over the unknown parameters $\lbrace \vect{\theta}_t, \beta_t, \vhat_t \rbrace$. The velocity $\vect{v}_{t,\mathrm{sup}}$ that maximizes Eq.~\ref{eq:likratio} can be computed explicitly
\begin{align}
	v_{t,\mathrm{sup}} &= \frac{|\vect{v}_0 + C\sigma_v^2\vect{\mathcal{T}} |}{1 + C^2 \sigma_v^2 \mathcal{N}^2 } \\
	\vhat_{t,\mathrm{sup}} \cdot \vhat_0 & = \frac{ C\sigma_v^2 \vect{\mathcal{T}}\cdot\vect{v}_0 + v_0^2}{v_0 |\vect{v}_0 + C\sigma_v^2\vect{\mathcal{T}} |}.
\end{align}
Using the above expression and dropping the constant terms in Eq.~\ref{eq:likratio}, we find the expression for the optimal test statistic

\begin{align}
\mathcal{R} & = \sup_{\vect{\theta}_t, \beta_t} \ln \frac{ \mathcal{L}\left( \lbrace \vect{\mu}_i \rbrace \big |M_l,r_l, f_l \right)}{\mathcal{L}\left( \lbrace \vect{\mu}_i \rbrace \big |\text{no lens} \right)} \Bigg|_{\vect{v}_t = \vect{v}_{t,\mathrm{sup}}} \label{eq:test_statistic_R2} 
\end{align}
given by the expression in Eq.~\ref{eq:test_statistic_R}.

\subsection{Raw data}
\label{subsec:raw_data}

In this part of the Appendix, we describe our data manipulations in more depth.

\begin{figure}[tbp]
	\includegraphics[width=.4\textwidth,origin=0,angle=0]{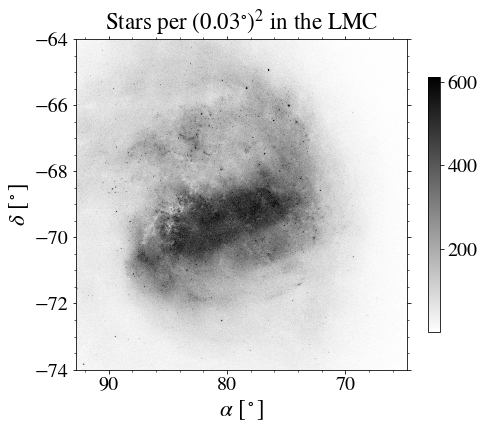}
	\includegraphics[width=.4\textwidth,origin=0,angle=0]{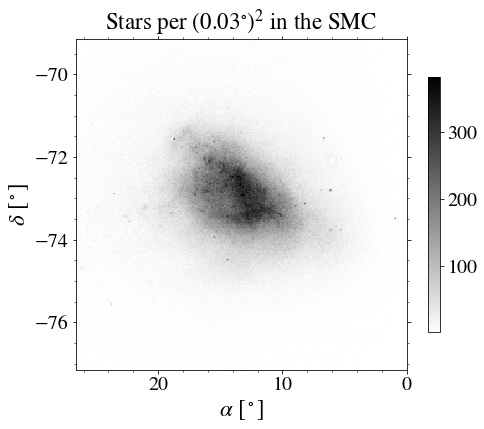}
	\caption{Number density distribution on the sky of the stars selected from the Large Magellanic Cloud (top) and Small Magellanic Cloud (bottom).} \label{fig:star_count}
\end{figure}

\subsubsection{Data Cleaning }
\label{p:subtraction}

In Fig.~\ref{fig:star_count}, we plot the stellar surface densities of the Magellanic Clouds (MCs), as they appear in the second data release of {\it Gaia} \cite{prusti2016gaia, brown2018gaia} for the selected 14,017,189 stars of the Large Magellanic Cloud (LMC) and the 1,890,713 stars of the Small Magellanic Cloud (SMC). 
Before applying the template analysis to the chosen stellar targets, we address systematics whose potential contributions to the proper motions of the stars could be misconstrued as a lensing signal. One source of contamination comes from overdense stellar clusters, which we remove by comparing a pixelated angular number density map $\Sigma(\vect{\theta})$ in pixels of size $\beta_{sm}/3$ with an average local number density map, smoothed with a Gaussian distance kernel of size $\beta_{\rm{sm}} = 0.1 ^\circ$:

\begin{align}
\label{eq:clumps}
\Sigma_{\rm{sm}}(\vect{\theta}) =  \frac{\int \dd\vect{\theta'} \exp\left[-\frac{(\vect{\theta} - \vect{\theta'})^2}{2\beta_{\rm{sm}}^2}\right]  \Sigma(\vect{\theta'})}{\int \dd\vect{\theta'} \exp\left[-\frac{(\vect{\theta} - \vect{\theta'})^2}{2\beta_{\rm{sm}}^2}\right]}.
\end{align}
We remove pixels for which $\Sigma(\vect{\theta}) > 3\Sigma_{\rm{sm}}(\vect{\theta})$, reducing the initial LMC and SMC star populations by $0.03 \%$ and $0.10 \%$, respectively.

\begin{figure}[tbp]
	\begin{center}
		\includegraphics[width=0.48\textwidth]{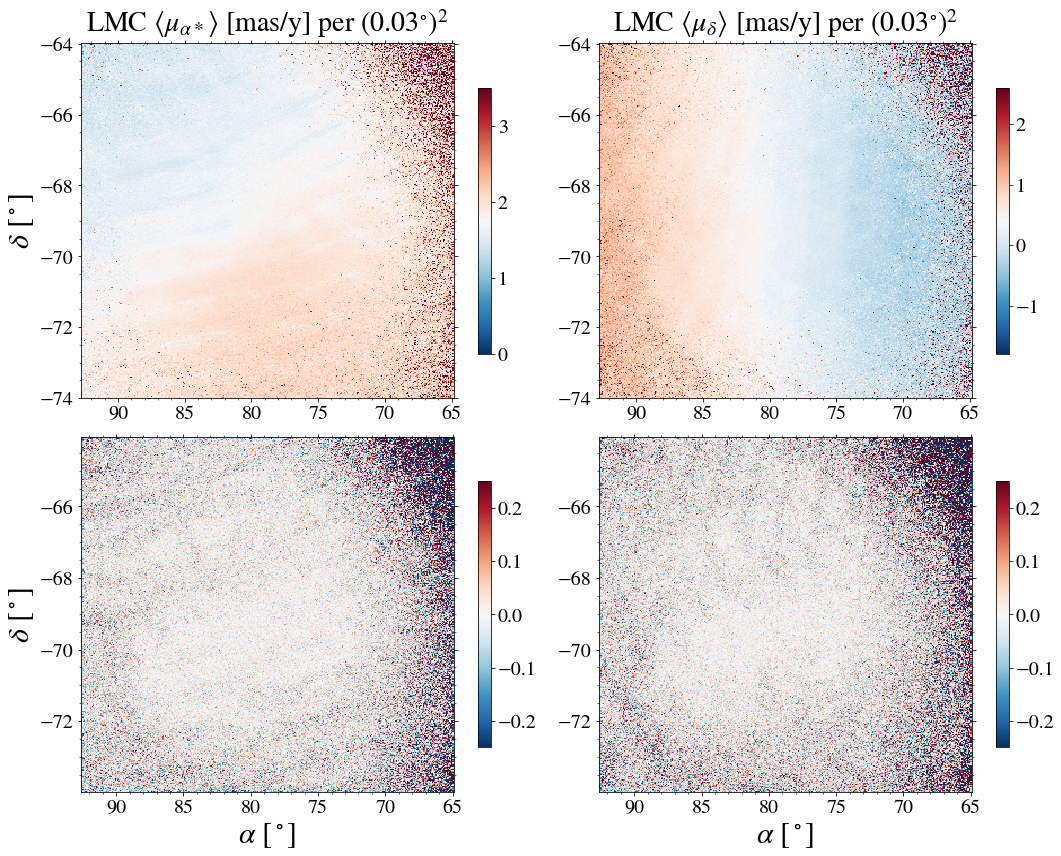}
		\caption[Proper motion of the stars in the Large Magellanic Cloud.]{Average stellar proper motion across the LMC, $\langle \mu \rangle \equiv \sum_{i \in p} \mu_i\sigma^{-2}_{\mu,i}/ \sum_{i \in p} \sigma^{-2}_{\mu,i}$ per pixels of size $0.03^\circ$ in the RA (left) and DEC (right) directions. The top panel shows the proper motion field in the original {\it Gaia} data sample after the removal of dense clusters, and the bottom panel after additional background motion subtraction and removal of outliers.} \label{fig:star_pm}
	\end{center}
\end{figure}

To account for the coherent velocity fields present in the data, we define a smoothed average local proper motion field $\vect{\mu}_{\rm{sm}}(\vect{\theta})$, again with a Gaussian kernel of angular size $\beta_{\rm{sm}} $

\begin{align}
\label{eq:v_subtraction}
\vect{\mu}_{\rm{\rm{sm}}}(\vect{\theta}) = \frac{\int \dd\vect{\theta'} \exp\left[-\frac{(\vect{\theta} - \vect{\theta'})^2}{2\beta_{\rm{sm}}^2}\right] \vect{\mu}(\vect{\theta'})}{\int \dd\vect{\theta'}  \exp\left[-\frac{(\vect{\theta} - \vect{\theta'})^2}{2\beta_{\rm{sm}}^2}\right]},
\end{align}
and subtract it from the local proper motion field $\vect{\mu}(\vect{\theta}_p) = \sum_{i \in p} \vect{\mu}_i\sigma_{\mu, i}^{-2}/\sum_{i\in p} \sigma_{\mu, i}^{-2}$, 
in square pixels of size $0.05^{\circ}$, to obtain $\vect{\mu}_{\mathrm{sub},i} \equiv \vect{\mu}_i - \vect{\mu}_\mathrm{sm}(\vect{\theta}_i)$. 
We cut out ``velocity outliers''---gravitationally unbound, high-velocity stars---that do not satisfy the relation $\mu_{\mathrm{sub},i} > 3 \sigma_{\mu,i} + \mu_\mathrm{esc}$, where we take $\mu_{\rm{esc}} = 0.2$ mas/y as the (proper) escape velocity for both of the MCs. 
Removing velocity outliers causes the local mean proper motion to shift away from zero again. For this reason, we repeat the background motion subtraction and removal of outliers for a total of 3 iterations. The fraction of outliers in the last iteration is small enough to guarantee a final sample whose mean proper motion is consistent with zero. The velocity subtraction by means of a Gaussian distance kernel introduces edge artifacts of size $\sim \beta_{\rm{sm}}$ at each iteration, which we avoid by rejecting stars within $6\times \beta_{\rm{sm}}$ from the edges. Table~\ref{tab:percentage_dec} summarizes the fraction of stars removed from the original {\it Gaia} sample during each cleaning procedure. 
In Fig.~\ref{fig:star_pm}, we present the proper motion distribution of the LMC data with dense clusters removed, before (top panels) and after (bottom panels) the large-scale motion subtraction and removal of velocity outliers.

\begin{table}
	\begin{center}
		\begin{tabular}{| p{4cm}|p{1.3cm}|p{1.3cm}|p{3cm} }
			\multicolumn{3}{c}{Number of stars removed} \\
			\hline
			Cleaning procedure & LMC & SMC \\
			\hline
			dense clusters & $0.03 \%$& $0.10 \%$ \\
			velocity outliers, iter 1 & $ 5.32 \%$ & $0.66 \%$\\
			velocity outliers, iter 2 & $ 0.05 \%$ & $< 0.01 \%$\\ 
			velocity outliers, iter 3 & $ < 0.01 \%$ & $< 0.01 \%$\\
			edges & $6.91\%$ & $2.79 \%$\\
			\hline
		\end{tabular}
		\caption{Percentage decrease in the selected stellar populations due to the cleaning procedures described in the text.}	
		\label{tab:percentage_dec}
	\end{center}
\end{table}

\subsubsection{Calculation of the effective error}
\begin{center}
	\begin{figure}[tbp]
		\includegraphics[width=0.45\textwidth]{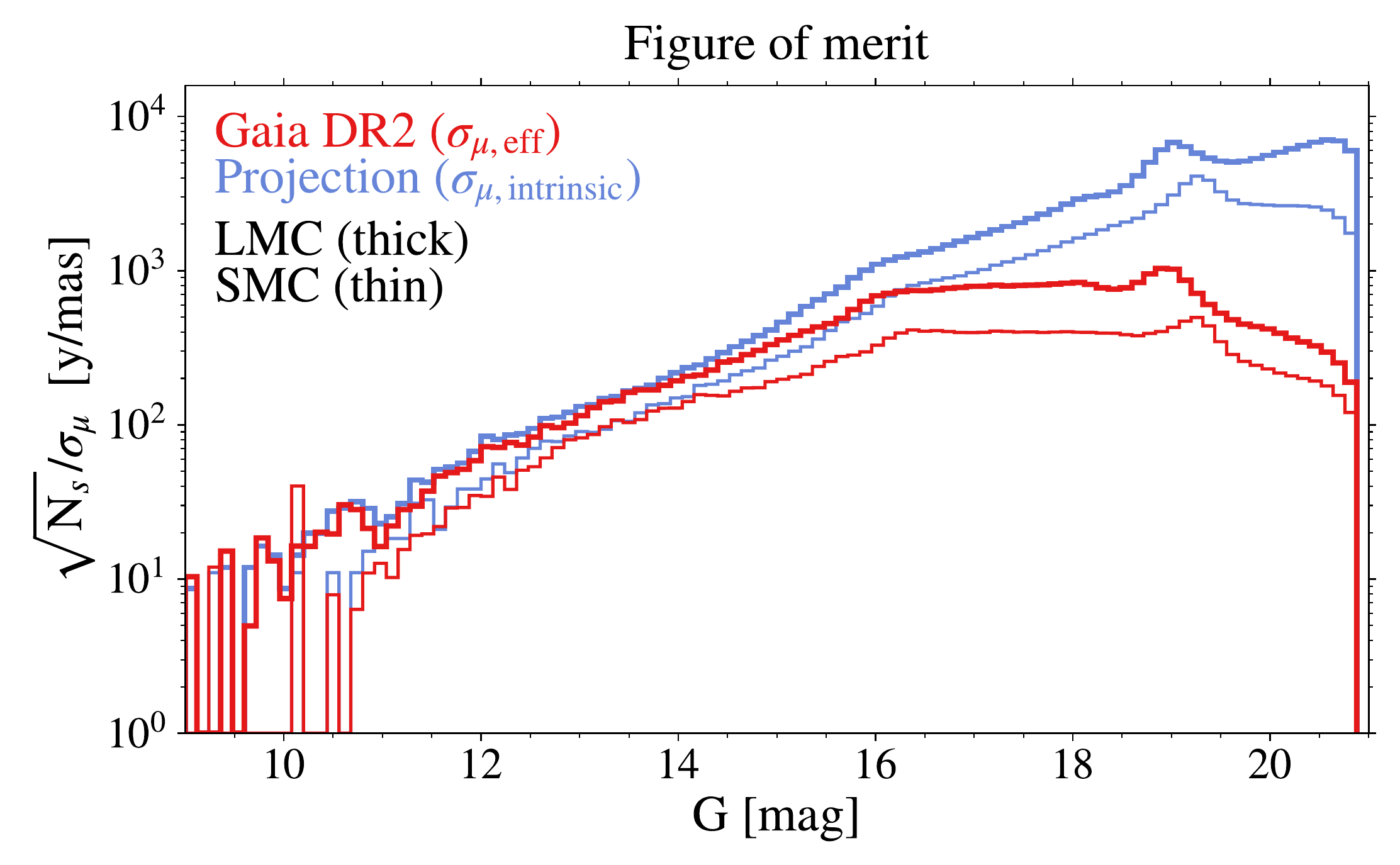}
		\caption{Figure of merit for the template velocity observable $\sqrt{N_s}/\sigma_\mu$ as a function of G-magnitude in the LMC (thick) and SMC (thin). $N_s$ is the number of stars per G-magnitude bin and $\sigma_\mu$ defined as in Fig.~\ref{fig:pm_disp}. The best sensitivity for the signal is given by the stars that maximize the figure of merit, i.e. with $16 \lesssim G \lesssim 19$ in {\it Gaia} DR2 (red) and potentially by the most numerous fainter stars when the measurements will reach the intrinsic proper motion dispersion floor (blue).}\label{fig:FOM}
	\end{figure}
\end{center}

This part of our data treatment deals with the spread and uncertainties in the stellar proper motions.
In general, the astrometric performance of $\it{Gaia}$ can be impacted by noise contributions both stochastic and systematic in nature. Even if we are in no position to precisely identify all the potential noise sources, the best way to acknowledge their presence is by calculating the mean effective variance $\sigma^{2}_{\rm{\mu, eff}}$. Since the precision of the astrometric measurements depends on the apparent brightness of the sources, we bin the stars in G-magnitude and calculate the effective dispersion in each bin. This quantity virtually encapsulates the instrumental errors, intrinsic dispersion, as well as artifacts of crowding, unresolved binaries, and potential residual trends in the stellar motions that were not successfully removed during the velocity subtraction. 
In each bin of $0.1$ in G-magnitude we contrast the mean observed effective error, $\sigma^{2}_{\rm{\mu, eff}} \equiv \langle \mu_{\alpha *}^2 + \mu_{\delta}^2  \rangle$, to the {\it Gaia}-reported formal error of each star, $\sigma_{\mu_i}^2 \equiv \sigma_{\mu_{i, \alpha}}^2 + \sigma_{\mu_{i, \delta}}^2$. In the vast majority of cases, the former exceeds the latter; in the computation of the test statistic $\mathcal{T}$, we weigh each star by the larger of the two quantities.  
We can deduce which stars have the best sensitivity for the lensing signal from the signal-to-noise ratio estimate of Eq.~\ref{eq:SNRlocal}. The relevant figure of merit to be maximized is $\sqrt{\Sigma}/\sigma_\mu$, where $\Sigma \propto N_s$ is the typical angular number density of a population of $N_s$ stars in the observed patch of the sky. As displayed in Fig.~\ref{fig:FOM}, the best sensitivity is currently coming from the population of stars with $16 \lesssim G \lesssim 19$. 
Though the $\it Gaia$ DR2 catalog showcases significant advancements compared to DR1, its capabilities in terms of astrometric lensing searches are still far from their ultimate end-of-mision values due to the relatively short observational period. This current dataset still contains partial instrumental calibration errors, inadequate background estimation, underestimates of centroid location uncertainties, and mislabeling of the sources' properties, among other unmodelled errors~\cite{arenou2018gaia, lindegren2018gaia}. 
The improvement of such issues in view of the increased time span of the operational phase, along with the scaling of the proper motion error with time as $ t_{\rm{int}}^{-3/2}$, are bound to drive up the value of the figure of merit in the near future.

\subsection{Template method details}
\begin{center}
	\begin{figure}[tbp]
		\includegraphics[width=0.45\textwidth]{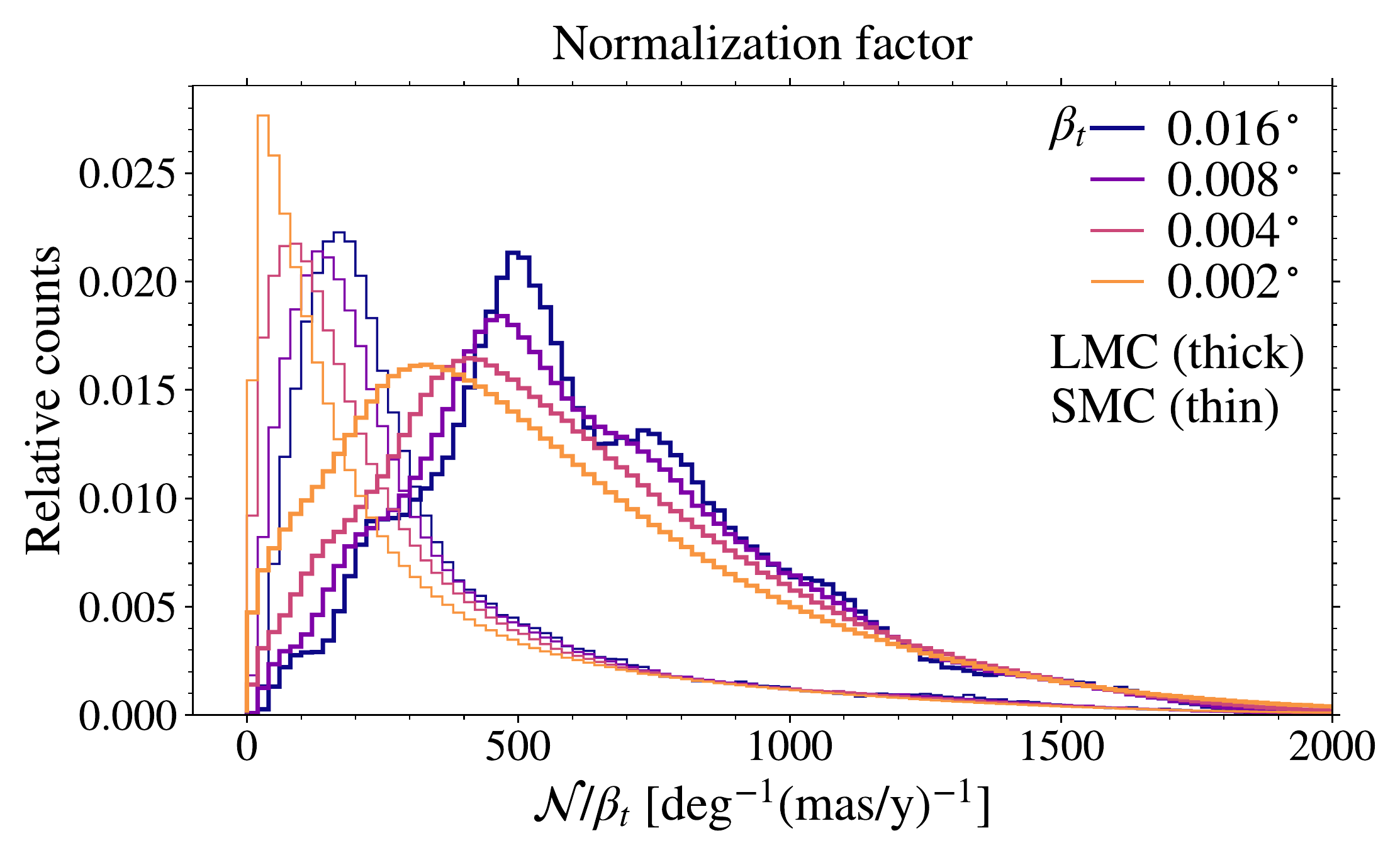}
		\caption{Histograms of the normalization factor defined in Eq.~\ref{eq:template_norm_simp} at selected angular scales $\beta_t$ for the LMC (thick) and the SMC (thin). The expected scaling $\mathcal{N}/\beta_t \sim \sqrt{\Sigma}/\sigma_\mu$ is observed in the data for $\beta_t \gtrsim 0.004^\circ$, and breaks down at scales smaller than the typical angular separation between two stars.}\label{fig:Nplot}
	\end{figure}
\end{center}

To perform the template scanning at a given angular scale $\beta_t$, the stars are pixelated with pixel size $0.1\times \beta_t$. We first scan coarsely every 9 lattice sites, computing $\mathcal{T}$ as defined in Eq.~\ref{eq:template_stat}, along the horizontal ($\vhat_t = \vect{\hat{\alpha}}$) and vertical ($\vhat_t = \vect{\hat{\delta}}$) directions, and $\mathcal{N}$ as defined in Eq.~\ref{eq:template_norm_simp}, using a masking matrix kernel of size $\beta_\mathrm{mask} = \mathrm{max}(4\times \beta_t, 0.01^\circ)$. The normalization factors obtained from the MCs scanning at 4 different angular scales are displayed in Fig.~\ref{fig:Nplot}. For a uniformly distributed set of background sources with angular number density $\Sigma$, we expect the distribution and typical value of $\mathcal{N}/\beta_t \sim \sqrt{\Sigma}/\sigma_\mu$ to be independent of $\beta_t$, a behavior borne out in the data for $\beta_t \gtrsim 0.004^\circ$. The expected scaling breaks down for $\beta_t$ values smaller than the typical angular separation between two stars, which in practice sets a lower bound on the useful angular scales to be considered for the template.
For each parameter space point $\lbrace M_l, r_l, f_l\rbrace$ we compute the optimal template angular scale $\beta_{t, \rm{opt}}$, defined by requiring 3 expected lenses in front of the stellar targets with $\beta_l \geq \beta_{t, \rm{opt}}$. We then select from the fixed list of 58 $\beta_t$ values the 3 closest to $\beta_{t,\rm{opt}}$ (or the 2 closest if $\beta_{t, \rm{opt}} \leq 0.0015^\circ$ or $\beta_{t, \rm{opt}} \geq 0.03^\circ$). This subset is used to perform the template scanning on a fine grid with lattice constant $\beta_\mathrm{scan} = 0.1\times \beta_t$ on squares of size $\beta_t$ centered at the location of the simulated lenses (only the 200 closest lenses are retained). The list of $\lbrace \mathcal{T}_\alpha, \mathcal{T}_\delta, \mathcal{N} \rbrace$ values obtained is used to compute the test statistic $\mathcal{R}$ to be compared with the value resulting from the coarse+fine scan of both the LMC and SMC data using the same subset of $\beta_t$ values. For computational efficiency, we only compute the test statistic in locations where $C\sigma_v\mathcal{N} > 1$, as we do not expect a large signal elsewhere.

\subsection{Lens density profiles}
\begin{figure}[tbp]
	\includegraphics[width=0.48\textwidth]{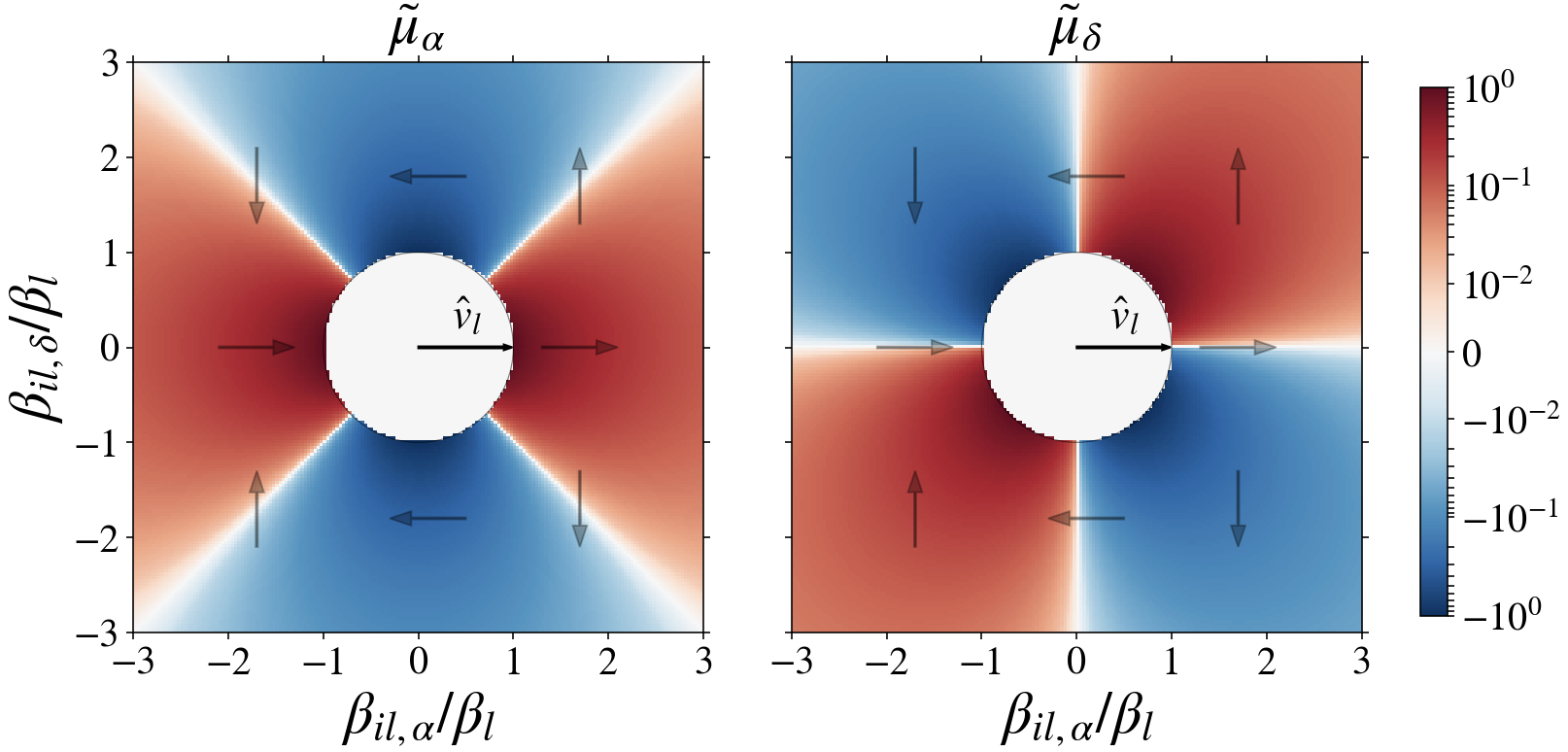}
	\caption{Right-ascension ($\alpha$, left) and declination ($\delta$, right) components of the angular velocity vector profile $\tilde{\vect{\mu}}_i$ of Eq.~\ref{eq:mupattern} as a function of angular separation $\vect{\beta}_{il}$ taking $\widetilde{M}_l(\beta_{il}) = \Theta(\beta_{il} - \beta_l)$ and $\partial_{\beta_{il}} \widetilde{M}(\beta_{il}) = 0$. The lens has angular size $\beta_l$ and is moving in the direction $\vhat_l = \vhat_\alpha$.}\label{fig:dipoleTrunc}
\end{figure}

The analysis presented in the main text is repeated using a velocity template that ignores the details of the inner density profile of the lens by truncating the signal at angular distances $> \beta_l$. In Eq.~\ref{eq:mupattern} we take $\widetilde{M}_l(\beta_{il}) = \Theta(\beta_{il} - \beta_l)$ and $\partial_{\beta_{il}} \widetilde{M}(\beta_{il}) = 0$, obtaining the universal dipole pattern displayed in Fig.~\ref{fig:dipoleTrunc}. The resulting limit on compact lenses from the MCs data analysis is presented in Fig.~\ref{fig:compactlimit_tr} in the $(M_l, f_l)$ plane for three values of $r_l$. The result is comparable to the limit shown in Fig.~\ref{fig:compactlimit} obtained using the lens profile of Eq.~\ref{eq:lensprofile}. Note that in the presence of a lens with an unknown profile, the analysis using the truncated velocity template would still capture a large fraction of the signal. The sensitivity is expected to improve by an $\mathcal{O}(1)$ factor when choosing the velocity template that exactly matches the distortion produced by the true lens.

\begin{figure}[tbp]
	\includegraphics[width=0.45\textwidth]{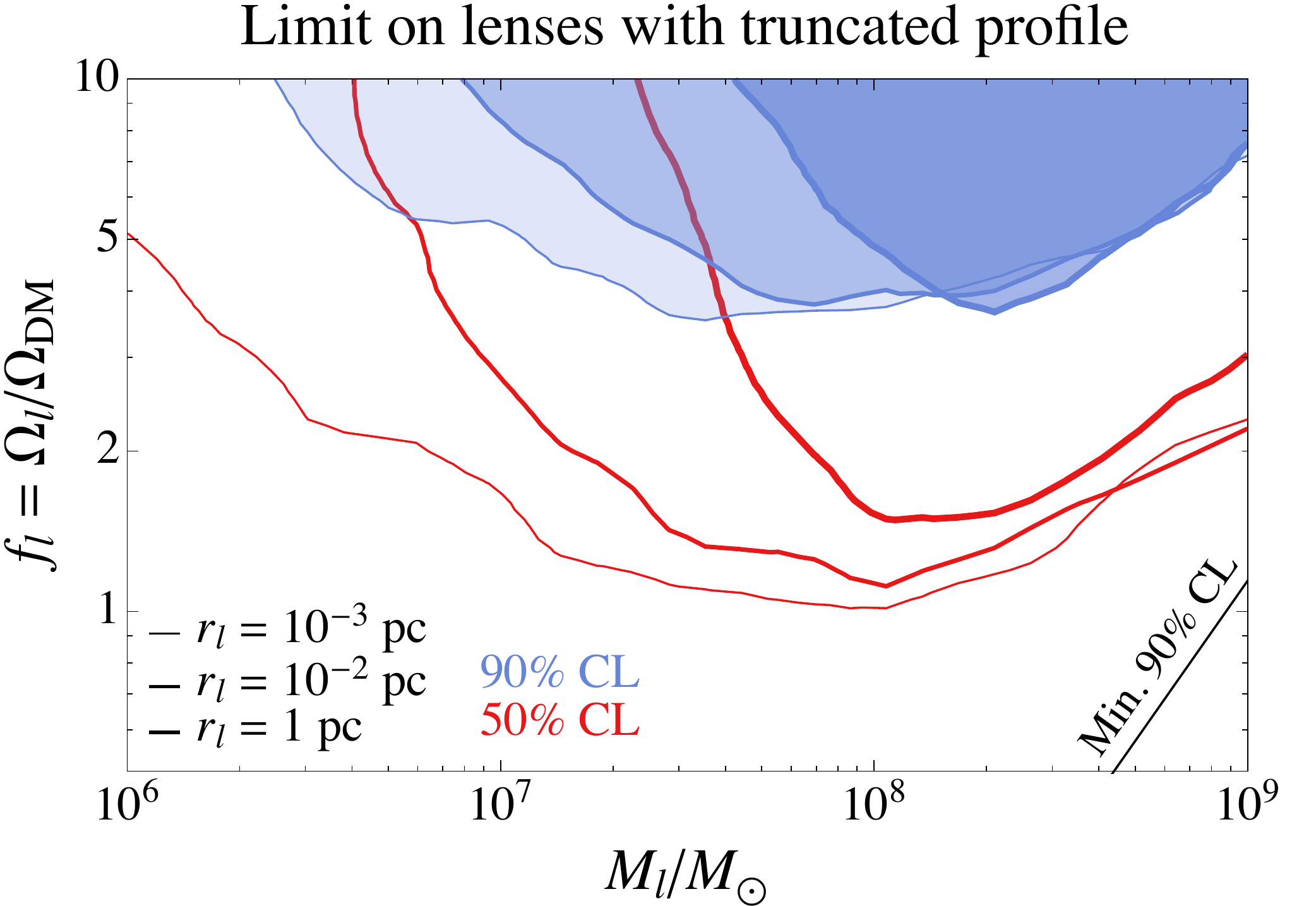}
\caption{Constraints from the MCs velocity template analysis on the fractional dark matter abundance $f_l$ of compact objects with mass $M_l$ and density profile  truncated at $r_l$, for different compact object radii $r_l = 10^{-3}, 10^{-2},\text{ and }1~\text{pc}$. The constraint for the smallest radius is equivalent to the one for point-like objects ($r_l = 2 G_N M_l$) given the angular number density of stars. Above the diagonal line at the bottom right, at least one subhalo eclipses the data sample with 90\% probability. }
\end{figure}

\bibliography{lensvelocity}

\end{document}